\def\noi{\noindent}
\def\be{\begin{equation}}
\def\ee{\end{equation}}
\def\bea{\begin{eqnarray}}
\def\eea{\end{eqnarray}}
\def\nn{\nonumber}
\def\um{\frac{1}{2}}
\def\uc{\frac{1}{4}}
\def\tr{{\rm Tr}}
\def\d{{\rm d}}

\documentclass[11pt]{article}

\usepackage{color}
\usepackage{amsfonts}

\begin{document}

\begin{center}
{\Large {\bf A Quantizable Model of Massive Gauge Vector Bosons\\ without Higgs}}
\end{center}

\bigskip
\bigskip

\centerline{V. Aldaya$^{1,a}$, M. Calixto$^{b,a}$ and F.F.
Lopez-Ruiz$^{a}$}

\bigskip
\centerline{\it $^{a}$ Instituto de Astrof\'\i sica de Andalucia
(IAA-CSIC), Apartado Postal 3004, Granada 18080, Spain.}
\centerline{\it $^{b}$ Departamento de Matem\'atica Aplicada y
Estad\'\i stica, Universidad Polit\'ecnica de Cartagena,}
\centerline{\it Paseo Alfonso XIII 56, 30203 Cartagena, Spain.}

\footnotetext[1]{Corresponding author. E-mail address: valdaya@iaa.es (V. Aldaya).}

\bigskip

\begin{center}

{\bf Abstract}
\end{center}

\small
\setlength{\baselineskip}{12pt}

\begin{list}{}{\setlength{\leftmargin}{3pc}\setlength{\rightmargin}{3pc}}
\item We incorporate the  parameters of the gauge group $G$ into
the gauge theory of interactions through a non-linear
partial-trace $\sigma$-model Lagrangian on $G/H$. The minimal
coupling of the new (Goldstone-like) scalar bosons provides mass
terms to those intermediate vector bosons associated with the
quotient $G/H$, without spoiling gauge invariance, remaining the
$H$-vector potentials massless. The main virtue of a partial trace
on $G/H$, rather than on the entire $G$, is that we can find an
infinite-dimensional symmetry, with non-trivial Noether
invariants, which ensures quantum integrability in a non-canonical
quantization scheme. The present formalism is explicitly applied
to the case $G=SU(2)\times U(1)$, as a Higgs-less alternative to
the Standard Model of electroweak interactions, although it can
also be used in low-energy phenomenological models for strong
interactions.
\end{list}

\normalsize

\noi PACS: 12.15.-y, 12.60.-i, 11.15.-q, 02.20.Tw, 12.10.-g.
\setlength{\baselineskip}{14pt}


\section{Introduction}

In the 1960's the mechanism of spontaneously broken symmetry,
usually referred to as the Higgs-Kibble mechanism \cite{Higgs},
came into the particle physics scenario \cite{Weinberg}, imported
from solid state physics (mainly in relation to Meissner effect),
to match the masses of the intermediate vector bosons with
renormalizability \cite{'tHooft}. However, in spite of the wide
acceptance today of the Standard Model of electroweak interactions
as a whole and of its phenomenological accuracy (putting aside the
existence of the Higgs particle), there exists a rather extended
feeling that a deeper structure is underneath,  owing specially to
the artificiality of the mass generation mechanism.

In this paper we face the chief point of the mass generation
mechanism aiming at outlining a conceptually and mathematically
neat framework within which the fundamentals of the Standard Model
can be reproduced. This framework is essentially based on the
inclusion of the gauge-group parameters into the theory as scalar
dynamical fields paralleling the standard Goldstone bosons.
With a proper Lagrangian for these new fields of the
$\sigma$-model type and appropriate rewriting of the traditional
Minimal Coupling Prescription we arrive at a general Massive Gauge
Theory explicitly exhibiting gauge symmetry. When applied to the
electroweak symmetry the new prescription provides mass to the
$W^{(\pm)}$ and $Z$ vector bosons without the need for the Higgs
particle, leaving naturally the electromagnetic field massless. It
might also be used to address low energy effective models for the
strong interaction according to the schemes handled in Refs.
\cite{Japoneses}.

The explicit use of the Goldstone bosons in the description of
physical processes is by no means new in the literature. In fact,
as a consequence of the widely named ``Equivalence Theorem''
\cite{Logitano,Bardeen}, according to which a very heavy Higgs
particle can be eliminated from the broken symmetry programme in
favour of non-linear $\sigma$-like Goldstone bosons, the actual
computation of Feynman diagrams involving the longitudinal
polarizations of the (massive) vector bosons in electroweak
interactions can be resolved in terms of the corresponding
diagrams among those scalar fields. But even more, the possibility
of incorporating explicitly the Goldstone bosons into the theory as
some sort of matter fields has also been considered in the
framework of non-abelian (generalized) Stueckelberg theory without
Higgs \cite{Kunigoto}. Unfortunately, the use of a non-linear
$\sigma$-Lagrangian, as a trace over the whole gauge group, has
led to an insoluble dichotomy unitarity-renormalizability
\cite{dicotomia,Hurth} (see also the review \cite{StueckelbergRev}
and references therein).

In this paper we introduce a simple, though essential,
modification to the non-abelian Stueckelberg model. We shall adopt
a non-linear partial-trace $\sigma$-model Lagrangian on $G/H$
instead of on the whole $G$. The minimal coupling of the new
(Goldstone-like) scalar bosons provides mass terms to those
intermediate vector bosons associated with the quotient $G/H$,
without spoiling gauge invariance, so that the $H$-vector
potentials remain massless in a natural way. The advantage of considering
a partial trace on $G/H$, rather than on the entire $G$, lies on
the existence of an infinite-dimensional symmetry enlarging the
gauge symmetry group, providing as many non-zero Noether
invariants as field degrees of freedom in the solution manifold of
the physical system. This ensures quantum integrability, at least
under a non-canonical quantization scheme based on symmetry
grounds, as has been widely demonstrated in those systems bearing
enough symmetries as happens in, for instance, conformal field
theories.

It is well known that the non-linear sigma model, in general,
suffers from unavoidable renormalizability problems under the
canonical quantization programme (see, for instance,
\cite{Ketov}). In fact, the trouble that canonical quantization
faces in dealing with systems bearing non-trivial topology could
be traced back to the ``tangent space" approximation imposed at
the very beginning of the (canonical) quantization program
\cite{Isham}. Already in the simple case of ``free" particles
moving on spheres, a proper quantization requires the replacement
of canonical commutators with the Lie-algebra commutators of the
Euclidean group \cite{Isham,sigmita}. Going further in this
direction, we shall replace canonical commutators between
coordinates and momenta with Lie-algebra commutators between group
generators of the enlarged local symmetry. In fact, the new
``canonical'' structure of the solution manifold can be derived
directly from the symmetry group as one of its canonical invariant
forms (giving the symplectic potential).

The present paper is organized as follows. In Sec. 2 we briefly
report on the partial-trace version of the non-Abelian
Stueckelberg Model. In Sec. 3, a finite quantization of the
$G=SU(2)$ case will be sketched, although our symmetry-based
quantization proposal can be applied as well to any semi-simple
gauge group. In Sec. 4 we consider the case of $G=SU(2)\times
U(1)$, aiming at reformulating the fundamentals of the Standard
Model for electroweak interactions, along with an additional
simple symmetry mixing  which generates the mass of the elementary
charged fermions.

\section{A revision of the non-Abelian Stueckelberg model}

As already commented in the introduction, Stueckelberg's original
idea (versus Proca) for giving mass to $U(1)$ gauge vector bosons
in a renormalizable way consists roughly speaking in giving
dynamical content to the gauge group parameters
$U(x)=e^{i\varphi(x)}\in U(1)(M), x\in M$, through the kinematical
$U(1)-\sigma$ model Lagrangian (density)
\be \mathcal{L}^{U(1)}_\sigma=\um \partial_\mu U \partial^\mu
U^{\dag}=\um
\partial_\mu \varphi\partial^\mu\varphi=\theta_\mu\theta^\mu,
\ee
($\theta_\mu\equiv-i\partial_\mu U U^\dag=\partial_\mu\varphi$)
minimally coupled according to the standard prescription:
\be \tilde{\mathcal{L}}^{U(1)}_\sigma=\um (D_\mu U)(D^\mu
U)^{\dag}=\um (\theta_\mu-A_\mu)(\theta^\mu-A^\mu)\ee
where $D_\mu=\partial_\mu-iA_\mu$ stands for the covariant
derivative. Then, the complete Abelian Stueckelberg Lagrangian is:
\be {\mathcal{L}}^{U(1)}_{\rm MA}=-\uc
F^{\mu\nu}F_{\mu\nu}+m^2\tilde{\mathcal{L}}^{U(1)}_\sigma.\ee
Quite remarkably, this Lagrangian is gauge invariant, even though
it explicitly contains mass terms for the gauge bosons and no
symmetry breaking has taken place.

The natural non-Abelian extension of the Stueckelberg formalism
for a general gauge group $G$ follows similar steps. Now
$U(x)=e^{i\varphi^a(x)T_a}\in G(M)$, where $T_a, a=1,\dots,{\rm
dim}(G)$ are the Lie-algebra generators of $G$ with commutation
relations $[T_a,T_b]=C_{ab}^cT_c$. We shall restrict ourselves to
unitary groups and set the normalization
$\tr(T_aT_b)=\delta_{ab}$. When referring to the canonical 1-form
on $G$, we must distinguish between the left- and right-invariant
ones: $\theta_\mu^L=-iU^\dag\partial_\mu U$ and
 $\theta_\mu\equiv\theta_\mu^R=-i\partial_\mu U U^\dag$,
 respectively. The $G$-invariant $\sigma$-model Lagrangian now reads:
\be \mathcal{L}_\sigma^{G}=\um \tr(\partial_\mu U \partial^\mu
U^{\dag})=\um\tr(\theta_\mu\theta^\mu)=\um\tr(\theta^L_\mu\theta^{L\mu})\equiv\um
g_{ab}(\varphi)\partial_\mu\varphi^a\partial^\mu\varphi^b \ee
which is highly non-linear and chiral. The minimal coupling is
formally analogous to the Abelian case, namely
\be \tilde{\mathcal{L}}_\sigma^{G}=\um \tr((D_\mu U)(D^\mu
U)^{\dag})=\um \tr(
(\theta_\mu-A_\mu)(\theta^\mu-A^\mu)),\label{tlGs}\ee
although $A_\mu$ must be understood as $A_\mu=A_\mu^aT_a$. This
Lagrangian is invariant, in particular, under
\be U\to VU, A_\mu\to VA_\mu V^\dag-i\partial_\mu V
V^\dag.\label{gaugetrans}\ee
Adding the standard kinematical Lagrangian for Yang-Mills fields
${\mathcal{L}}_{\rm YM}^G=-\uc \tr(F^{\mu\nu}F_{\mu\nu})$, with

\be F_{\mu\nu}(A)\equiv
\partial_{\mu}A_{\nu}-
\partial_{\nu}A_{\mu}+
[A_{\mu},A_{\nu}]\label{fea},\ee
to (\ref{tlGs}), we arrive at the full Lagrangian for Massive
Yang-Mills bosons
\be {\mathcal{L}}_{\rm MYM}^{G}={\mathcal{L}}_{\rm
YM}^G+m^2\tilde{\mathcal{L}}_\sigma^{G}.\label{chocho}\ee
As already mentioned in the introduction, this model prevents the
massive Yang-Mills theory from being both unitary and
renormalizable, at least in the canonical quantization approach.

Our main proposal here in this paper lies on a revision of this
model consisting in restricting the whole trace on $G$ to a
partial trace on a quotient manifold $G/H$. $H$ is the isotropy
subgroup of a given direction $\lambda=\lambda^aT_a$, in the
Lie-algebra of $G$, under the adjoint action $\lambda\to V\lambda
V^\dag$, where $\lambda^a$ are real numbers subjected to
$\tr(\lambda^2)=1$. Let us define $\Lambda\equiv U\lambda U^\dag$.
The claimed $G/H-\sigma$ Lagrangian has the following expression:
\be \mathcal{L}_\sigma^{G/H}=\um \tr([-iU^\dag\partial_\mu
U,\lambda]^2)\equiv\um\tr([\theta_\mu^L,\lambda]^2)=\um\tr([\theta_\mu,\Lambda]^2)
=\um\tr((\partial_\mu\Lambda)^2).
\label{trazapar}
\ee
The minimally coupled version
\be \tilde{\mathcal{L}}_\sigma^{G/H}=\um \tr([-iU^\dag D_\mu
U,\lambda]^2)=\um\tr([\theta_\mu-A_\mu,\Lambda]^2) \ee
is again gauge invariant under (\ref{gaugetrans}). As in
(\ref{chocho}), the partial-trace ($G/H$) Massive Yang-Mills
Lagrangian now follows:

\be {\mathcal{L}}_{\rm MYM}^{G/H}={\mathcal{L}}_{\rm
YM}^G+m^2\tilde{\mathcal{L}}_\sigma^{G/H}.\label{ltot1}\ee
We should remark that the change of variables
\be \tilde{A}_\mu=U^\dag(A_\mu-\theta_\mu)U=U^\dag A_\mu U+i
U^\dag\partial_\mu U\label{acurva}\ee
and the fact that $F(A)=U F(\tilde{A}) U^\dag$, renders the
Lagrangian (\ref{ltot1}) into the simple form
\be {\mathcal{L}}_{\rm MYM}^{G/H}=-\uc
\tr(F^{\mu\nu}(\tilde{A})^2)+\um
m^2\tr([\tilde{A}_\mu,\lambda]^2). \label{ltot2}\ee
This change of variables, formally mimicking the shift to the
unitary gauge, turns the actual degrees of freedom of the theory
apparent. On the other hand, it must be eventually completed with
the change of variables $\phi=U^\dag\psi$ when the fermionic
matter field $\psi$ will be introduced.

For instance, for $G=SU(2)$ with the standard spherical basis
$T_\pm,T_0$ of the Lie algebra, and taking $\lambda=\lambda^0T_0$,
that is $H=U(1)$, the mass term in (\ref{ltot2}) is written
\be \um m^2\tr([\tilde{A}_\mu,\lambda]^2)=m_{W}^2
\tilde{W}^{+}_\mu\tilde{W}^{\mu -},\ee
where, as usual, $\tilde{W}^\pm_\mu=\tilde{A}_\mu^1\pm
i\tilde{A}_\mu^2$ and $m_{W}=m\lambda^0$, and
$\tilde{W}^0_\mu=\tilde{A}^3_\mu$ remains massless.

\section{Quantization}
For linear systems, the quantum theory proceeds according to the
usual canonical quantization rules, which lead to commutators
between basic operators realizing the Lie algebra of the
Heisenberg-Weyl group in the corresponding dimension. In field
theories, this means postulating equal-time commutation relations
between fields and their time derivatives, or conjugate momenta,
$[\phi(x), \pi(y)] = i \delta(x-y)$. Going to nonlinear systems
with non-flat phase space should require a different approach.
This is precisely the situation we are facing now, as a result of
the introduction of the group parameters as physical degrees of
freedom, with a (curved) compact target space $G/H$.

Our strategy is to look for a replacement of the Heisenberg-Weyl
group with a (more involved) symmetry group of the solution
manifold, keeping the general idea of considering as basic
conjugate operators those giving central terms under commutation.
Therefore, we should be able to identify such symmetry group by
analyzing the symplectic potential (or Liouville 1-form) in the
solution manifold, which generalizes $p_i \d q^i$ from particle
mechanics.\footnote{The symplectic potential can be obtained by
integrating the Lagrangian Poincar\'e-Cartan form on a Cauchy
hypersurface $\Sigma$.} Regarding the sigma sector, and from
(\ref{trazapar}), this can be written:
\bea \Theta_\sigma^{G/H}&=&\int_\Sigma\tr([\theta_\mu,\Lambda]
[-i\delta U U^{\dag},\Lambda])\d \sigma^\mu \label{thetagrupo}\\
&=&\int_\Sigma \tr(\{[\theta_\mu,\Lambda] + \Lambda \tr(\Lambda
\theta_\mu)\} [-i\delta U U^{\dag},\Lambda])\d \sigma^\mu \equiv
\int_\Sigma \tr( \vartheta_\mu[-i\delta U U^{\dag},\Lambda])\d
\sigma^\mu ,\nn \eea
where in the second step we have added a null contribution
intended to make the change
\be \theta_\mu\to \vartheta_\mu \equiv [\theta_\mu,\Lambda] +
\Lambda \tr(\Lambda \theta_\mu), \label{cambiorbita} \ee
invertible. The key observation is that the group of
transformations (with parameters $U'$ and $\vartheta'_\mu$)
\bea
U &\rightarrow& U' U \nn\\
\vartheta_\mu & \rightarrow& U' \vartheta_\mu U'^{\dag} +
\vartheta'_\mu \label{euclideo}\eea
\noindent renders (\ref{thetagrupo}) invariant up to a total
differential. This can be easily checked, taking into account the
identity $\delta \Lambda \equiv  \delta (U\lambda U^{\dag}) =
[\delta U U^{\dag}, U\lambda U^{\dag}]$.

It should be noted that the choice of coordinates
(\ref{cambiorbita}) also represents a deviation from the canonical
prescription; it generalizes to field theory the selection of
angular momentum $\vec{L}$ and position $\vec{q}$ as basic
conjugate variables for the quantum mechanical motion on the
sphere $ {\mathbb S}^2$ \cite{sigmita}. A similar change to (\ref{cambiorbita})
applies to the vector potentials,
\be \mathcal A_\mu \equiv [A_\mu,\Lambda] + \Lambda \tr(\Lambda
A_\mu), \label{cambiorbitaa} \ee
which now acquire the same composition law as in the second line in
(\ref{euclideo}), without being forced to be a ``flat connection''.
Once we combine (\ref{cambiorbita}) and  (\ref{cambiorbitaa})
together with (\ref{acurva}), we can define
\be \tilde \mathcal A_\mu = U^{\dag}(\mathcal A_\mu
-\vartheta_\mu) U = [\tilde A_\mu,\lambda] + \lambda \tr(\lambda
\tilde A_\mu). \ee
The complete symplectic potential for the Massive Yang-Mills
theory is written in terms of the new $\tilde \mathcal A_\mu$ as
\be \Theta_{MYM}^{G/H}=\int_\Sigma \tr(F^{\mu\nu}(\tilde \mathcal
A)\delta\tilde \mathcal A_\nu + m^2 \tilde \mathcal A^\mu
[-i\delta U U^{\dag},\lambda])\d \sigma_\mu. \label{toalateta} \ee
This expression tells us directly the actual conjugate couples of
basic ``coordinates'' and ``momenta'' to be quantized. It is
apparent now that the components of $\tilde \mathcal A _\mu$
perpendicular to $\Sigma$ and $\lambda$ (in space-time and group
directions, respectively) have the gauge group parameters  $U$
themselves as conjugate coordinates.

Let us go back to the original variables $A_\mu^a$ and write
\be \hat{G}_a(x)=\frac{\delta}{\delta
\varphi^a(x)}+C_{ab}^c\varphi^b(x)\frac{\delta}{\delta
\varphi^c(x)}+\dots\ee
for the generators of gauge transformations, and denote
$\hat{E}_a^\mu(x)$ for the generator of translations in
$A_\mu^a(x)$ and $\hat{A}_i^a(x)$ for the generator of
translations in $F^{0i}_a(x)\equiv E^i_a(x)$, in much the same way
the momentum $\hat{p}$ is the generator of translations in $q$ in
standard Quantum Mechanics.

Choosing $\Sigma= {\mathbb R}^3$ in the time direction (i.e., $\d
\sigma_\mu\to\d^3x$), we propose the following equal-time
($x^0=y^0$) commutators:
\bea \left[\hat{G}_a(x),\hat{G}_b(y)\right] &=& iC_{ab}^c
\hat{G}_c(x)\delta(x-y),\nn\\
\left[\hat{A}_j^a(x),\hat{E}^k_b(y)\right] &=&
i\delta_j^k\delta^a_b\delta(x-y),\label{ym-com}
\\ \left[\hat{G}_a(x),\hat{A}^b_j(y)\right] &=& iC_{ac}^b
\hat{A}^c_j(x)\delta(x-y)
-i\delta_a^b\partial_j^x\delta(x-y),\nn\\
 \left[\hat{G}_a(x),\hat{E}_b^\mu(y)\right] &=& iC_{ab}^c \hat{E}_c^\mu(x)\delta(x-y)
 -im^2\delta^\mu_0C_{ab}^c\lambda_c\delta(x-y)\nn,\eea
as corresponds to the Lie-algebra of the symmetry group of the
system (see  Appendix). The commutator in the last line of
(\ref{ym-com}) algebraically expresses the comment after
(\ref{toalateta}) concerning the conjugated character of
translations in $A_0$ and $U$.

 A unitary, irreducible representation
of this (infinite-dimensional) Lie algebra on wave functionals
$\Psi(E^\mu)$ in the ``electric-field representation'' $E^\mu_a$,
where $E^0_a\equiv m^2\tr(T_a\Lambda)$, can be achieved as (the
actual details will be given in \cite{largo}):
\bea \hat{E}_a^\mu\Psi(E)&=&
(E_a^\mu-m^2\delta^\mu_0\lambda_a)\Psi(E),\nn\\
\hat{A}^a_j\Psi(E)&=& i\frac{\delta}{\delta E_a^j}\Psi(E),\nn\\
\hat{G}_a\Psi(E)&=& \left(\vec{\nabla}\cdot \vec{E}_a +iC_{ab}^c
(E_c^\mu-m^2\delta^\mu_0\lambda_c)\frac{\delta}{\delta
E_b^\mu}\right)\Psi(E)
 \eea
The last expression accounts for the non-Abelian ``Gauss law" when
the constraint condition $\hat{G}_a\Psi(E)=0$ is required.

 It should be stressed
that the central term proportional to $\lambda_c$ in the last
commutator of (\ref{ym-com}) could also be considered as a remnant
of some sort of ``symmetry breaking'' in the sense that it can be
hidden into a redefinition of $\hat{E}_a^0$,
\be \hat{E}_a^0\rightarrow
\hat{E}'^0_a\equiv\hat{E}_a^0+m^2\lambda_a,\label{shiftE} \ee
which now acquires a non-zero vacuum expectation value
proportional to the mass $m^2\lambda_a$, that is:
\be\langle 0|\hat{E}_a^0|0\rangle=0\longrightarrow \langle
0|\hat{E}_a^0|0\rangle= m^2\lambda_a.\ee

\section{The $SU(2)\times U(1)$ group and the Standard Model}

 In this section
we shall denote by $B_\mu=B^a_\mu T_a, a=1,\dots,4$, the
$SU(2)\times U(1)$ Lie-algebra valued vector
potential, keeping $A_\mu$ for the electromagnetic potential, as
usual. The new generator $T_4$ ($\equiv\um Y$, the halved
hypercharge) corresponds to the direct factor $U(1)$.

The key point in this section consists in combining the
construction above for $G=SU(2)$ with the traditional Stueckelberg
model for a selected $H^\perp= U(1)$. However, this time we shall
choose $\lambda$ in the electric charge (mixed) direction
\be \lambda\propto Q\equiv T_3+T_4,\label{GM}\ee
according to the usual Gell-Mann-Nishijima relation, and we shall
choose $H^\perp$ in the orthogonal direction:
\be Q^\perp=T_3-T_4,\label{GMp}\ee
in the sense that $\tr(QQ^\perp)=0$. This way we shall provide
mass to three vector bosons, say $W^\pm_\mu\propto B^1_\mu\pm
iB^2_\mu$ and $Z_\mu\propto \tr(Q^\perp B_\mu)$, out of the
original four vector potentials $B_\mu^a, a=1,\dots,4$, leaving
the electromagnetic potential $A_\mu$ massless. In fact, the
Standard Model Lagrangian for the Yang-Mills sector will be
\bea {\cal L}_{\rm MYM}^{\rm SM}&=&-\uc \tr (F_{\mu\nu})^2+\um
{m^2}\tr([\theta_\mu-B_\mu,UQU^\dag]^2)+\um
{m'^2}\tr(((\theta_\mu-B_\mu)UQ^\perp U^\dag)^2)\nn\\
&=&-\uc \tr (F_{\mu\nu})^2+\um {m^2}\tr([\tilde{B}_\mu,Q]^2)+\um
{m'^2}\tr((\tilde{
B}_\mu Q^\perp)^2) \\
&\equiv&-\uc F^a_{\mu\nu}F_a^{\mu\nu}+m_{W}^2
\tilde{W}^{+}_\mu\tilde{W}^{\mu -}+\um
m^2_{Z}{\tilde{Z}}_\mu{\tilde{Z}}^\mu, \eea
where $\tilde{B}_\mu$ is related to $B_\mu$ in a way similar to
that of Eq. (\ref{acurva}). This Lagrangian reproduces the
Yang-Mills sector of the Standard Model for electroweak
interactions when we introduce the usual coupling constants
$g,g',e$ according to $\tilde{B}^3_\mu\equiv g{\cal B}^3_\mu,
\tilde{B}^4_\mu\equiv g'{\cal B}^4_\mu,
\tilde{Z}_\mu\equiv\frac{gg'}{e}{\cal Z}_\mu$. Writing ${\cal
Z}_\mu$ in terms of ${\cal B}^3_\mu$ and ${\cal B}^4_\mu$, we
have:
\be {\cal Z}_\mu=\frac{e}{gg'}\tilde{ Z}_\mu=\frac{e}{gg'}(g{\cal
B}^3_\mu-g'{\cal B}^4_\mu) \equiv\cos(\vartheta_W){\cal
B}^3_\mu-\sin(\vartheta_W){\cal B}^4_\mu,\ee
which, together with the orthogonal relation
\be {\cal A}_\mu\equiv\sin(\vartheta_W){\cal
B}^3_\mu+\cos(\vartheta_W){\cal B}^4_\mu,\ee
(the electromagnetic vector potential) defines the usual Weinberg
rotation of angle $\vartheta_W$.

\subsection{Giving mass to fermionic matter}

The introduction of mass for fermionic matter can be accomplished
by a nontrivial mixing between space-time and internal symmetries.
Although the general setting of this symmetry mixing is rather
ambitious, here we shall consider the consequences of the
simplest, nontrivial, mixing between the Poincar\'e group
${\mathcal P}$ and the electromagnetic gauge subgroup $H=U(1)_Q$,
which has been widely developed in \cite{Eduardo} and references
therein. A more general symmetry mixing scheme, involving
conformal symmetry and larger internal symmetries, is under
consideration \cite{conformalmixing}.

To be precise, we propose a mass-generating mechanism associated
with a non-trivial mixing of  the Poincar\'e group  and
$SU(2)\times U(1)$. This mixing takes place
through a linear combination ${P}'_0\equiv P_0+\kappa Q$ between
the time translation generator $P_0$ and $Q$, in much the same way
the generator $Q$ had to be found as a linear combination of $T_3$
and $T_4$. The spirit of the redefinition $P'_0$ is the same as
the shifting (\ref{shiftE}) [with $\lambda\propto Q$], ultimately
responsible for the mass $m_W$. In fact, with the new mass
operator
\be {M}'^2\equiv {{P}'_0}^2-\vec{P}'^2 \ee
the mass shell condition for fermionic fields $\psi$ becomes
\be {M}^2 \psi\,=({{P}_0}^2-\vec{P}^2)\psi=m_0^2\psi\to
\,{M}'^2\psi=\,(m_0^2+2\kappa P_0 Q+\kappa^2Q^2)\psi. \ee
At the rest frame we have
\be {M}'^2 \psi\,=\,(m_0^2+2\kappa m_0 Q+\kappa^2Q^2)\psi\;. \ee

\noi Then, for ``originally'' massless particles ($m_0=0$),
\be {M}'^2 \psi\,=\,\kappa^2Q^2\psi\;,\ee

\noi so that only charged particles acquire mass. This is in
agreement with the fact that there is no elementary fermionic
massive particles without electric charge.

\appendix
\section{Symmetry group of massive Yang-Mills theories\label{appendix}}
In this appendix we simply provide the composition law of the
infinite-dimensional symmetry group from which the physical
operators and their commutators (\ref{ym-com}) can be explicitly
derived as right-invariant vector fields. The group law:
\bea
U''(x)&=&U'(x)U(x) \ \ (\Rightarrow \  \theta''_\mu(x)= U'(x)\theta_\mu(x) U'^\dag(x)+\theta'_\mu(x)),\nn\\
A''_\mu(x)&=& U'(x)A_\mu(x) U'^\dag(x)+A'_\mu(x),\nn\\
F''_{\mu\nu}(x)&=& U'(x)F_{\mu\nu}(x) U'^\dag(x)+F'_{\mu\nu}(x),\\
\zeta''&=&\zeta'\zeta \exp\left(i\int_\Sigma
d\sigma^\mu(x)J_\mu(U',A',F';U,A,F)\right),\nn\\
J_\mu&=&J_\mu^{{\rm YM}}+J_\mu^\sigma,\nn\\
 J_\mu^{{\rm YM}}&=&\um\tr\left((A'^\nu-\theta'^\nu)U'F_{\mu\nu}U'^\dag-
 F'_{\mu\nu}U'(A^\nu-\theta^\nu)U'^\dag\right),\nn\\
J_\mu^\sigma&=&m^2\tr\left(\lambda(U'(A_\mu-\theta_\mu)U'^\dag-(A_\mu-\theta_\mu))\right),
\nn \eea
is a central extension by $U(1)\ni \zeta$ of the basic symmetry
group containing gauge transformations $U$ and translations in $A$
and $F$. This central extension is given by a two-cocycle
$\int_\Sigma d\sigma^\mu(x)J_\mu(U',A',F';U,A,F)$ defined through
a symplectic potential current $J_\mu$ made of two pieces: $
J_\mu^{{\rm YM}}$ accounting for the symplectic structure of the
pure Yang-Mills theory and $J_\mu^\sigma$ concerning the sigma
(massive) sector. The unitary, irreducible representations of this
infinite-dimensional symmetry group will be explicitly given in
\cite{largo}, inside a Group Approach to Quantization scheme.

\section*{Acknowledgements}

V. Aldaya and F.F. Lopez-Ruiz would like to thank T.W.B. Kibble
and H.F. Jones for reading the manuscript and for useful
discussions, and Imperial College and specially H.F. Jones for
their kind hospitality. We all thank A.P. Balachandran, J.
Guerrero, C. Barcel\'o and J.L. Jaramillo for encouraging
discussions. Work partially supported by the Fundaci\'on S\'eneca,
Spanish MICINN and Junta de Andaluc\'\i a under projects
08814/PI/08, FIS2008-06078-C03-01 and FQM219, respectively. F.F.
Lopez-Ruiz thanks C.S.I.C. for an I3P grant. M. Calixto thanks the
Spanish MICINN for a mobility grant PR2008-0218.


\end{document}